\documentstyle{mn}
\input{psfig}
\newcommand{\etal}{{et al.~}}

\newcommand{\lta}{\la}

\newcommand{\kms}{\>{\rm km}\,{\rm s}^{-1}}

\newcommand{\kpc}{\>{\rm kpc}}

\newcommand{\MLsun}{\>({\rm M}/{\rm L})_{\odot}}

\newcommand{\apj}{ApJ}

\newcommand{\aj}{AJ}
\newcommand{\mnras}{MNRAS}
\newcommand{\aap}{A\&A}

\newcommand{\nat}{Nature}

\newdimen\hssize
\newdimen\hdsize
\begin{document}


\title[The Angular Momentum Content of Dwarf Galaxies]
      {The Angular Momentum Content of Dwarf Galaxies:\\
       New Challenges for the Theory of Galaxy Formation}
\author[F.C. van den Bosch, A. Burkert and R.A. Swaters]
       {Frank C. van den Bosch$^{1}$, Andreas Burkert$^{2}$, and 
        Rob A. Swaters$^{3}$ \\
        $^1$Max-Planck Institut f\"ur Astrophysik, Karl Schwarzschild
         Str. 1, Postfach 1317, 85741 Garching, Germany\\
        $^2$Max-Planck Institut f\"ur Astronomie, K\"onigstuhl 17,
         D-69117 Heidelberg, Germany\\
        $^3$Carnegie Institution of Washington, Washington DC 20015, USA}


\date{}

\pagerange{\pageref{firstpage}--\pageref{lastpage}}
\pubyear{2000}

\maketitle

\label{firstpage}


\begin{abstract}
  We  compute the specific  angular momentum distributions of a sample
  of low  mass disk galaxies observed   by Swaters.  We  compare these
  distributions to  those of  dark matter  haloes obtained  by Bullock
  \etal  from  high  resolution $N$-body   simulations  of   structure
  formation in   a $\Lambda$CDM Universe.  We   find that although the
  disk  mass fractions are significantly   smaller than the  Universal
  baryon fraction, the total specific angular momenta of the disks are
  in good  agreement with those  of dark matter  haloes. This suggests
  that disks  form  out  of only  a  small fraction  of  the available
  baryons,  but yet  manage   to draw most   of the  available angular
  momentum.  In   addition   we  find  that  the    angular   momentum
  distributions of disks  are clearly distinct from  those of the dark
  matter; disks lack predominantly  both low and high specific angular
  momentum.  Understanding  these  findings  in  terms of  a  coherent
  picture for  disk formation  is challenging.  Cooling,  feedback and
  stripping, which are the main  mechanisms to explain the small  disk
  mass fractions found,  seem  unable  to simultaneously explain   the
  disk's angular  momentum distribution.  In  fact,  it seems that the
  baryons that  make up the disks must  have been  born out of angular
  momentum  distributions that  are  clearly  distinct  from those  of
  $\Lambda$CDM haloes.  However, the  dark and baryonic mass component
  experience the same tidal forces,  and it is therefore expected that
  they should have similar  angular momentum distributions. Therefore,
  understanding the angular  momentum content of disk galaxies remains
  an important challenge for our picture of galaxy formation.
\end{abstract}




\section{Introduction}
\label{sec:intro}

Disk  galaxies are rotationally  supported  systems whose structure is
governed by angular momentum.  In   the current paradigm of  structure
formation, galaxies form hierarchically by the assembly of dark matter
haloes and the subsequent cooling of the baryonic  mass component.  In
this standard picture, protogalaxies acquire angular momentum by tidal
interactions   with   neighboring      protogalaxies;   the  so-called
cosmological torques (Hoyle 1953).   This mechanism of spinning up the
dark and   baryonic  matter has been   studied  in great  detail using
numerical simulations (e.g., Barnes  \& Efstathiou 1987;  Warren \etal
1992; Zurek, Quinn \&  Salmon 1988; Sugerman, Summers \&  Kamionkowski
2000) and found   to  be in  good agreement   with linear tidal-torque
theory (e.g., Doroshkevich 1970; White 1984; Catelan \& Theuns 1996).

Following Peebles (1969) it  has become customary to  parameterize the
angular  momentum  of dark   matter  haloes by  the  dimensionless spin
parameter
\begin{equation}
\label{spinparam}
\lambda = {J \vert E \vert^{1/2} \over G M^{5/2}}
\end{equation}
where $J$, $E$,  and $M$ are  the  total angular momentum, energy  and
mass of   the   halo, and   $G$ is  the   gravitational constant.  The
distribution     of  $\lambda$ is  well described     by  a log-normal
distribution,
\begin{equation}
\label{spindistr}
p(\lambda){\rm d} \lambda = {1 \over \sigma_{\lambda} \sqrt{2 \pi}}
\exp\biggl(- {{\rm ln}^2(\lambda/\bar{\lambda}) \over 2
  \sigma^2_{\lambda}}\biggr) {{\rm d} \lambda \over \lambda},
\end{equation}
with $\bar{\lambda}  \simeq  0.06$ and $\sigma_{\lambda}   \simeq 0.5$
(e.g., Barnes  \&  Efstathiou 1987;  Ryden 1988;  Cole  \& Lacey 1996;
Warren  \etal 1992).    Starting with the   seminal paper  of Fall  \&
Efstathiou (1980), it was soon realized that  the size distribution of
disk galaxies can be explained  as originating from the spin parameter
distribution  if  the assumption is made  that  baryons conserve their
specific angular momentum when cooling to form luminous galaxies. This
concept has spawn  our   current picture for   the formation  of  disk
galaxies,  which has been   addressed by numerous studies.  Blumenthal
\etal  (1986) and   Flores   \etal (1993) investigated   how adiabatic
contraction  of dark matter  haloes impacts on  the rotation curves of
disk galaxies.  Kauffmann (1996) linked the formation of disk galaxies
within this framework to the evolution of damped Ly$\alpha$ absorption
systems.   Dalcanton, Spergel \&  Summers (1997) and  Mo, Mao \& White
(1998) investigated the  structural properties of disks, with emphasis
on the variance  induced  by the  $\lambda$-distribution.   Subsequent
studies  included recipes for  bulge  formation,  gas viscosity,  star
formation and/or feedback and investigated more detailed properties of
these model disk galaxies, such  as the Tully-Fisher relation, the gas
mass fractions, and the  origin of the  Hubble sequence (van den Bosch
1998,   2000, 2001; Jimenez \etal   1998;  Natarajan 1999; Heavens  \&
Jimenez 1999; van den Bosch \& Dalcanton  2000; Firmani \& Avila-Reese
2000; Avila-Reese  \&  Firmani 2000; Efstathiou   2000; Zhang \&  Wyse
2000; Buchalter, Jimenez \& Kamionkowski 2001).

The standard picture  of disk  formation  that has  emerged from these
studies has been remarkably successful in explaining a wide variety of
observational  properties of  disk  galaxies.  However,  two important
problems, both related to the angular momentum  of disk galaxies, have
come to light.  First of  all, detailed hydro-dynamical simulations of
this  process of disk  formation in a  cold dark matter (CDM) Universe
yield disks that are an order of magnitude too  small (Navarro \& Benz
1991; White \& Navarro 1993; Steinmetz \& Navarro 1999). This problem,
known as the angular  momentum  catastrophe, is  a consequence of  the
hierarchical formation of galaxies which causes the baryons to lose a
large  fraction of their  angular  momentum to  the dark  matter.  The
second  problem concerns the actual     density distribution of   disk
galaxies.   Under the   assumption   of  detailed  angular    momentum
conservation, this density distribution  is a direct reflection of the
distribution of specific angular momentum of the protogalaxy. Although
the distribution   of {\it total}  specific   angular momentum of dark
matter haloes  is well established (i.e., equation~[\ref{spindistr}]),
little  is  known about  the actual distribution  of  specific angular
momentum in each individual halo.  Therefore, prompted by the observed
surface  brightness profiles of  disk  galaxies, most  models  for the
formation of disk galaxies  have made the   {\it a priori}  assumption
that the   disks  that form  have   exponential density distributions.
Recently,   however,  Bullock \etal  (2000)    determined  the angular
momentum distributions of  individual dark  matter haloes, and  showed
that disks  that  form out  of  such distributions are  more centrally
concentrated than an exponential.  Van den Bosch (2001) has shown that
when  star formation, bulge   formation, and  feedback are  taken into
account, the resulting  stellar density distributions are inconsistent
with observations,   at  least for  the  low  surface  brightness disk
galaxies.  In addition, van den Bosch showed that the truncation radii
of the resulting disks are too small compared to observations.

From  the above discussion  it is clear  that our understanding of the
formation of disk galaxies is directly related to the angular momentum
distribution   (hereafter  AMD)  of  the baryonic    mass component of
protogalaxies.   In    this paper   we   use  the   observed   density
distributions and rotation curves of a  sample of 14 dwarf galaxies to
compute their  AMDs.  A  comparison of  these AMDs  with those of dark
matter haloes then provides important clues to the formation process of
disk galaxies.

This  paper is  organized  as follows.  In Section~\ref{sec:theory} we
discuss the angular momentum  distributions expected on grounds of the
standard       paradigm     of         galaxy      formation.       In
Section~\ref{sec:angmomdistr}  we  then  present  the angular momentum
distributions of  a sample of  14 dwarf galaxies.  The implications of
these distributions for our picture of  galaxy formation are discussed
in    Section~\ref{sec:galform}, and   we  summarize   our results  in
Section~\ref{sec:summ}.

Whenever the cosmological framework is important for our discussion we
adhere  to the currently popular  $\Lambda$CDM model  with $\Omega_0 =
0.3$, $\Omega_{\Lambda}=0.7$, $h=0.7$,  and  $\sigma_8 = 1.0$.    This
model yields a reasonable  fit to  the  current suite  of cosmological
constraints, including  high  redshift  supernovae  (Perlmutter  \etal
1998; Riess  \etal 1998; Garnavich  \etal 1998), the  cosmic microwave
background radiation (e.g., de Bernardis \etal 2000), and the observed
cluster  abundances (Eke,  Cole \&  Frenk 1996;  Bahcall \& Fan 1998).
For the baryon  density we adopt $\Omega_{\rm bar}  = 0.019 \, h^{-2}$
as suggested by the observations of primordial deuterium abundances by
Tytler \etal (1999).    With  the cosmological parameters  defined  as
above this implies a Universal  baryonic mass fraction of $f_{\rm bar}
= \Omega_{\rm bar}/\Omega_0 = 0.13$.

\section{Theoretical background}
\label{sec:theory}

Consider a virialized halo with  mass $M_{\rm vir}$ consisting of dark
and baryonic matter, and let $p(j) {\rm d}j$  indicate the fraction of
mass with specific  angular momentum between  $j$ and $j +  {\rm d}j$.
The total specific angular momentum is given by
\begin{equation}
\label{zeta}
j_{\rm tot} = j_{\rm max} \left[ 1 - \int_{0}^{1} m(l) {\rm d}l
\right] \equiv \zeta j_{\rm max}. 
\end{equation}
Here $j_{\rm max}$ is the maximum specific angular momentum of $p(j)$,
$l   =  j/j_{\rm max}$,   and    $m(j)$ is the  normalized  cumulative
distribution
\begin{equation}
\label{mofj}
m(j) = \int_{0}^{j} p(j) {\rm d}j.
\end{equation}
The dimensionless spin  parameter $\lambda$ can  be related to $j_{\rm
tot}$ as
\begin{equation}
\label{lamjtot}
\lambda =  {j_{\rm tot} \over \sqrt{2} \,  r_{\rm vir} V_{\rm vir}} \,
{\cal G}.
\end{equation}
Here $r_{\rm vir}$ is the virial radius, $V_{\rm vir} = \sqrt{G M_{\rm
vir} / r_{\rm vir}}$, and ${\cal G}$ is a geometrical factor.

It is customary to parameterize the density distribution of virialized
structures as
\begin{equation}
\label{rhodm}
\rho_{\rm vir}(r) = \rho_s \left( {r \over r_s}\right)^{-\gamma} 
\left( 1 + {r \over r_s}\right)^{\gamma-3}
\end{equation}
with $r_s$ a scale-length and $\rho_s  = \rho_{\rm vir}(r_s)$. For the
density  distribution of the form~(\ref{rhodm}),  which reduces to the
NFW profile (Navarro, Frenk \& White 1996, 1997) for $\gamma=1.0$, one
finds
\begin{equation}
\label{geom}
{\cal  G} = {\cal  G}(\gamma,c_{\rm  vir}) = \sqrt{h(c_{\rm vir})}  \,
f^{-1}(c_{\rm vir}),
\end{equation}
where $c_{\rm   vir}  = r_{\rm vir}/r_s$   is  the halo  concentration
parameter,
\begin{equation}
\label{fff}
f(x) = \int_{0}^{x} {\rm d}y \, y^{2-\gamma} (1+y)^{\gamma-3},
\end{equation}
and
\begin{equation}
\label{hhh}
h(x) = x \, \int_{0}^{x} {\rm d}y \, f(y) \, y^{1-\gamma} (1+y)^{\gamma-3}.
\end{equation}
\begin{figure*}
\centerline{\psfig{figure=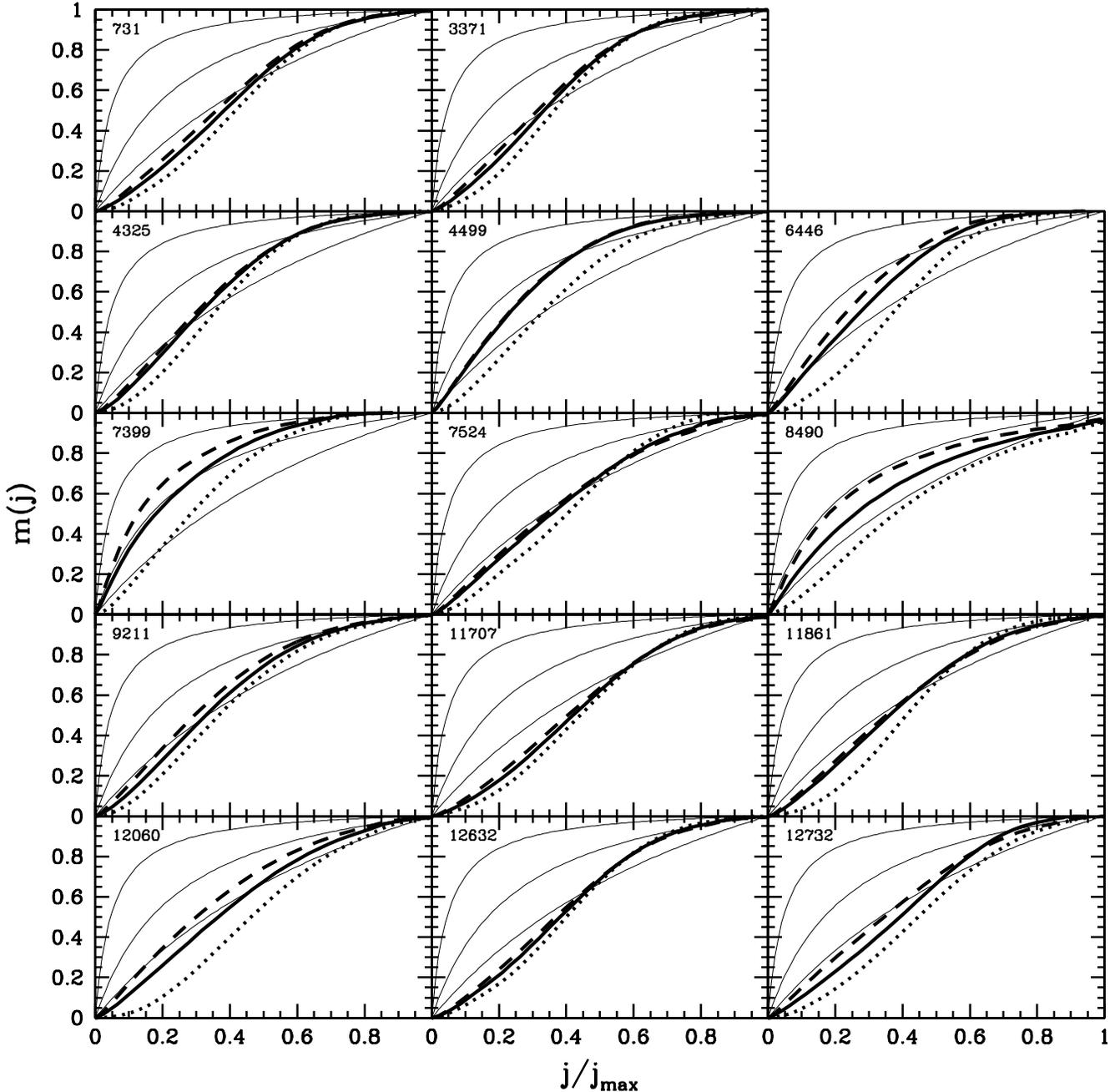,width=\hdsize}}
\caption{The   normalized  cumulative  angular momentum  distribution,
$m(j)$, as function of $j/j_{\rm max}$.  Each  panel plots, with thick
lines, three distributions    for   a particular   sample  galaxy  (as
indicated  in the panel) that  correspond to three different values of
the   stellar mass-to-light ratio:  $\Upsilon_R   = 0$ (dotted lines),
$\Upsilon_R = 1.0 \MLsun$ (solid  lines), and $\Upsilon_R=2.0  \MLsun$
(dashed   lines).   The three thin  lines   correspond to  AMDs of the
form~(\ref{mb00})  with  $\mu=1.06$ (upper curves), $\mu=1.25$ (middle
curves) and $\mu=2.0$  (lower curves) and  are plotted for comparison.
These values of $\mu$ correspond to the mean and  the 90 percent range
of  the distribution  in   $\mu$ found by  Bullock   \etal (2000)  for
$\Lambda$CDM  haloes. Note that  the majority of the  disks has an AMD
that is clearly distinct from that of the dark matter.}
\label{fig:allmj}
\end{figure*}

In the standard picture of disk formation it  is assumed that baryonic
and dark  matter experience the  same tidal torques,  and thus  end up
with the same  AMD $p(j)$. As shown by  Bullock \etal (2000, hereafter
B00) the AMD of dark matter haloes is well described by
\begin{equation}
\label{mb00}
m(l) = {\mu l \over l + \mu - 1},
\end{equation}
with  $l=j/j_{\rm  max}$.  For  dark matter  haloes  in a $\Lambda$CDM
Universe with  $\Omega_0 = 0.3$,  $\Omega_{\Lambda}=0.7$, $h=0.7$, and
$\sigma_8 = 1.0$, B00 found that the distribution of $\mu$ is Gaussian
in ${\rm log}(\mu - 1)$ with a mean of $-0.6$ and a standard deviation
of $0.4$. Thus for 90 percent of the haloes $1.06 < \mu  < 2.0$ with a
mean of $\langle \mu \rangle = 1.25$. For the probability distribution
corresponding to~(\ref{mb00}) one can write
\begin{equation}
\label{pb00}
p(s) = {\zeta \mu (\mu - 1) \over (\zeta s + \mu - 1)^2},
\end{equation}
with $s = j/j_{\rm tot}$ and
\begin{equation}
\label{zetab00}
\zeta = {j_{\rm tot} \over j_{\rm max}} = 
1 - \mu \left[ 1 - (\mu - 1) {\rm ln} \left(
{\mu \over \mu - 1}\right) \right].
\end{equation}
If our current picture for the  formation of disk galaxies is correct,
the baryonic  mass of  disk  galaxies thus should  have  an AMD of the
form~(\ref{pb00}).  However,  not necessarily  all the  baryons inside
$r_{\rm vir}$ are currently  part of the disk  (stars plus cold  gas).
Part of the baryons may not have cooled  by the present time, and/or a
certain fraction of  the baryons may have been  expelled from the halo
by either  feedback   processes  (i.e., galactic winds)  or    by some
stripping  mechanism.    We therefore define    the fraction   $f_{\rm
disk}=M_{\rm disk}/M_{\rm vir}$,  where $M_{\rm  disk}$ is the   total
disk mass observed. If  all  available baryons  have cooled to  become
part of  the disk one  expects that $f_{\rm disk} =  f_{\rm bar}$.  By
comparing  the   AMD   of the      observed   disk  with   that     of
equation~(\ref{pb00}), and by  investigating how  these AMDs correlate
with $f_{\rm disk}$, important insights  into the formation  mechanism
of dwarf  galaxies  can be  obtained.

\section{The angular momentum content of dwarf galaxies}
\label{sec:angmomdistr}

\subsection{Description of the data}
\label{sec:data}

In order to compute the angular  momentum distribution of the baryonic
component of a  disk  galaxy one needs  (i)  the rotation velocity  as
function of   radius, (ii) the  distribution of  the stellar mass, and
(iii) the distribution of the gas.  One thus requires a combination of
both  optical  and HI observations.   Since   most angular momentum is
contained by material at  large  galactocentric radii it is  essential
that the  observations probe  out to   large  enough radii.  Recently,
Swaters (1999) obtained both optical ($R$-band) and HI observations of
a large   sample of dwarf  galaxies,  down  to  fairly low  HI surface
brightness. These data are therefore ideally suited to compute angular
momentum   distributions.  A subset of  these  data,  with the highest
quality rotation  curves, was analyzed in  detail by van  den Bosch \&
Swaters (2001; hereafter BS01). Taking  account of the effects of beam
smearing and adiabatic contraction  they determined the  concentration
and the virial velocities of the dark  matter haloes, which they found
to be  consistent with  predictions of  $\Lambda$CDM  models.   In the
following we concentrate on the 14  galaxies for which BS01 achieved a
meaningful  fit.  These galaxies   are listed in Table~\ref{tab:data},
together with some global properties taken from Swaters (1999).
\begin{table}
\begin{minipage}{\hssize}
\caption{Properties of sample of late-type dwarf galaxies.}
\label{tab:data}
\begin{tabular}{rrccccc}
  UGC &  $D$   &  $M_R$   & $\mu_0^R$ & $R_d$ & $V_{\rm last}$ & $i$ \\
  731 &  $8.0$ & $-16.63$ & $23.0$ & $1.65$ & $74$ & $57$ \\
 3371 & $12.8$ & $-17.74$ & $23.3$ & $3.09$ & $86$ & $49$ \\
 4325 & $10.1$ & $-18.10$ & $21.6$ & $1.63$ & $92$ & $41$ \\
 4499 & $13.0$ & $-17.78$ & $21.5$ & $1.49$ & $74$ & $50$ \\
 6446 & $12.0$ & $-18.35$ & $21.4$ & $1.87$ & $80$ & $52$ \\
 7399 &  $8.4$ & $-17.12$ & $20.7$ & $0.79$ &$109$ & $55$ \\
 7524 &  $3.5$ & $-18.14$ & $22.2$ & $2.58$ & $79$ & $46$ \\
 8490 &  $4.9$ & $-17.28$ & $20.5$ & $0.66$ & $78$ & $50$ \\
 9211 & $12.6$ & $-16.21$ & $22.6$ & $1.32$ & $65$ & $44$ \\
11707 & $15.9$ & $-18.60$ & $23.1$ & $4.30$ &$100$ & $68$ \\
11861 & $25.1$ & $-20.79$ & $21.4$ & $6.06$ &$153$ & $50$ \\
12060 & $15.7$ & $-17.95$ & $21.6$ & $1.76$ & $74$ & $40$ \\
12632 &  $6.9$ & $-17.14$ & $23.5$ & $2.57$ & $76$ & $46$ \\
12732 & $13.2$ & $-18.01$ & $22.4$ & $2.21$ & $98$ & $39$ \\
\end{tabular}

\medskip

Column~(1) lists the  UGC number  of  the galaxy.  Columns~(2)  -- (7)
list the distance to the galaxy (in Mpc), absolute $R$-band magnitude,
central  $R$-band surface   brightness (in  mag arcsec$^{-2}$),  scale
length of  the stellar disk  (in kpc),  the observed rotation velocity
$V_{\rm last}$ (in $\kms$) at the last measured point, and the adopted
inclination angle (in  degrees), respectively.  Magnitudes and central
surface brightnesses have been corrected  for inclination and galactic
extinction,   but not for internal  extinction.  

\end{minipage}
\end{table}

\subsection{Mass modeling}
\label{sec:mass}

We now turn to computing the angular  momentum distributions of the 14
dwarf galaxies in our sample.   For that purpose  we first construct a
mass model of each  galaxy that best fits  the observed rotation curve
after the effects  of beam smearing  are  taken into account.  If  the
effects of  beam smearing are small,  one could  in principle directly
infer the  angular momentum  distribution  from the  observed  $V_{\rm
rot}(r)$ and the surface density  distributions of the stars and  gas.
However, the  advantage of the  mass modeling is  that typically HI is
measured out to much larger radii than the maximum radius out to which
$V_{\rm rot}$  is   determined   (simply because   a  measurement   of
$\Sigma_{\rm HI}$ requires a lower S/N than $V_{\rm rot}$).  Since low
surface brightness  HI at large radii  can still contain a significant
fraction of the total mass, and has high specific angular momentum, it
is  essential that one takes  this  gas into  account.   A proper mass
model that fits the observed  rotation curve makes predictions for the
circular velocity at all radii, and can thus be used to compute a more
complete  angular   momentum  distribution.   Furthermore,   the  mass
modeling  allows to correct for  the effects of  beam smearing.
\begin{figure*}
\centerline{\psfig{figure=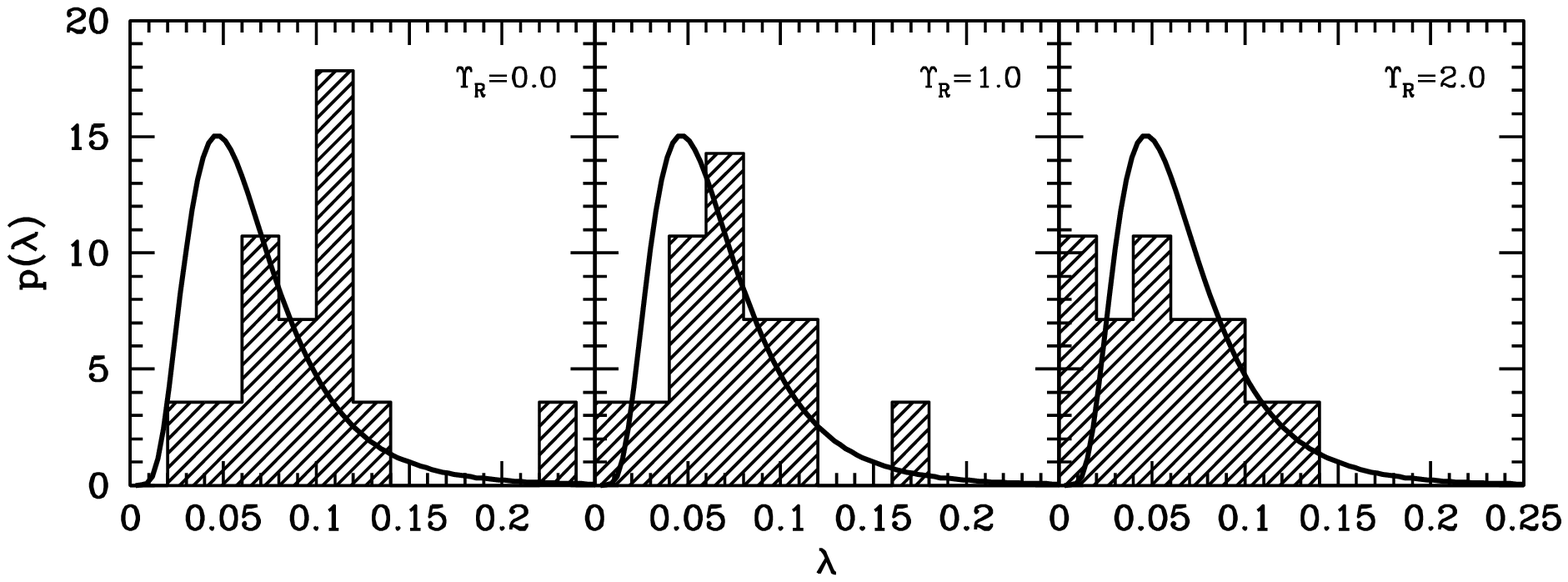,width=\hdsize}}
\caption{Histograms  (hatched)  of the  distribution  of $\lambda_{\rm
disk}$ for the 14 dwarf galaxies in our sample. These values have been
determined  from the   AMDs plotted  in  Figure~\ref{fig:allmj}  using
equations~(\ref{zeta})   and~(\ref{lambar}).  Results  are plotted for
three values of the stellar  mass-to-light ratio, as indicated in each
panel.  The thick  solid lines  plots the  probability distribution of
equation~(\ref{spindistr})   with    $\bar{\lambda}    =    0.06$  and
$\sigma_{\lambda} =   0.5$, and  is to  represent   the spin parameter
distribution  of   dark matter  haloes.    Especially for $\Upsilon_R =
1.0\MLsun$  the  two distributions  agree remarkably  well, suggesting
that the disks have the same  distribution of total specific angular
momentum as the dark matter haloes.}
\label{fig:histo}
\end{figure*}

For  our mass modeling we  use  the same  method as  employed in BS01,
which  we briefly repeat here  for completeness.  We assume that there
are three mass components in  each galaxy: an infinitesimally thin gas
disk, a thick stellar disk, and a spherical dark matter halo.

We make the assumption that the gas is distributed axisymmetrically in
an infinitesimally thin disk,  and model  the HI density  distribution
as:
\begin{equation}
\label{sbHI}
\Sigma_{\rm HI}(R) = \Sigma_0 \left({R \over R_1}\right)^{\beta}
{\rm e}^{-R/R_1} + f \; \Sigma_0 \, {\rm e}^{-((R-R_2)/\sigma)^2}.
\end{equation}
The first term represents an exponential  disk with scale length $R_1$
and with a central hole, the extent of  which depends on $\beta$.  The
second term corresponds  to a  Gaussian ring  with radius $R_2$  and a
FWHM $\propto \sigma$.  The flux ratio between these two components is
set  by  $f$.   The form  of   equation~(\ref{sbHI}) has no particular
physical motivation, but   is an appropriate  fitting  function, which
yields   excellent  fits   to  the   observed   HI surface  brightness
distributions (see BS01)\footnote{In  the case of UGC~7524 the fitting
function of equation~(\ref{sbHI}) can  not satisfactorily describe the
data  and  linear interpolation  between the data   points  is used to
describe  $\Sigma_{\rm   HI}(R)$.}.   When   computing   the  circular
velocities of the  gas, we multiply $\Sigma_{\rm  HI}$ by a factor 1.3
to correct for the contribution of Helium.
\begin{table}
\begin{minipage}{\hssize}
\caption{Results for $\Upsilon_R=1.0 \MLsun$ and $\gamma = 1.0$.}
\label{tab:results}
\begin{tabular}{rrrcccrr}
 UGC &  $c_{\rm vir}$   &  $V_{\rm vir}$   & $f_{\rm disk}$ & $f_{\rm
  gas}$ & $\lambda_{\rm disk}$ & $j_{\rm tot}$ & $j_{\rm max}$ \\
  731 & $17.6$ & $ 48.6$ & $0.024$ & $0.801$ & $0.061$ & $ 308$ & $ 775$ \\ 
 3371 & $10.6$ & $ 65.5$ & $0.018$ & $0.715$ & $0.056$ & $ 569$ & $1618$ \\ 
 4325 & $33.0$ & $ 48.6$ & $0.037$ & $0.541$ & $0.074$ & $ 328$ & $ 971$ \\ 
 4499 & $ 2.4$ & $126.3$ & $0.003$ & $0.672$ & $0.007$ & $ 330$ & $1195$ \\ 
 6446 & $ 9.1$ & $ 56.2$ & $0.041$ & $0.570$ & $0.052$ & $ 397$ & $1325$ \\ 
 7399 & $19.9$ & $ 65.8$ & $0.012$ & $0.691$ & $0.044$ & $ 396$ & $1692$ \\ 
 7524 & $ 6.4$ & $ 78.8$ & $0.012$ & $0.500$ & $0.025$ & $ 393$ & $1025$ \\ 
 8490 & $17.5$ & $ 53.2$ & $0.026$ & $0.769$ & $0.062$ & $ 378$ & $1106$ \\ 
 9211 & $19.2$ & $ 41.2$ & $0.055$ & $0.865$ & $0.107$ & $ 381$ & $1058$ \\ 
11707 & $14.6$ & $ 62.2$ & $0.062$ & $0.770$ & $0.103$ & $ 886$ & $2046$ \\ 
11861 & $16.4$ & $ 93.1$ & $0.068$ & $0.405$ & $0.099$ & $1861$ & $4820$ \\ 
12060 & $31.1$ & $ 42.8$ & $0.102$ & $0.710$ & $0.168$ & $ 582$ & $1477$ \\ 
12632 & $16.5$ & $ 47.8$ & $0.033$ & $0.760$ & $0.078$ & $ 387$ & $ 976$ \\ 
12732 & $ 9.0$ & $ 68.9$ & $0.040$ & $0.869$ & $0.081$ & $ 926$ & $2267$ \\ 

\end{tabular}

\medskip

Column~(1) lists  the UGC number of  the  galaxy.  Columns~(2) and~(3)
list $c_{\rm vir}$ and $V_{\rm vir}$ (in  $\kms$) of the best-fit mass
model, respectively.  Columns~(4)  and~(5) list the disk mass fraction
$f_{\rm disk}$ and the gas  mass fraction $f_{\rm gas}$, respectively.
Column~(6) lists the baryonic spin parameter $\lambda_{\rm disk}$, and
columns~(7)  and~(8),  list the   total and  maximum  specific angular
momentum of the baryons (both in $\kpc \kms$).

\end{minipage}
\end{table}

For the stellar disk we  assume a thick exponential 
\begin{equation}
\label{rhostar}   
\rho^{*}(R,z) = \rho^{*}_0  \,  \exp(-R/R_d) \, {\rm sech}^2(z/z_0) 
\end{equation} 
where  $R_d$  is  the  scale  length of  the   disk in   the $R$-band.
Throughout we set $z_0   = R_d/6$.  The  exact  value of  this  ratio,
however,  does not significantly influence   the results.  None of the
galaxies in our sample has a significant bulge component.

For   the  DM component  we consider   a   density distribution of the
form~(\ref{rhodm}).  Unless stated  otherwise we focus  on haloes with
$\gamma=1.0$ (i.e., NFW  profiles).  Rather than parameterizing the DM
halo by $(c_{200},V_{200})$, as  in BS01, we use $(c_{\rm  vir},V_{\rm
vir})$. Here
\begin{equation}
\label{cvir}
c_{\rm vir} = {r_{\rm   vir} \over r_s} = c_{200}  \biggl({\Delta_{\rm
vir} \over 200}\biggr)^{-1/3},
\end{equation}
with   $\Delta_{\rm vir}$ the virial   density, defined as the average
density  inside $r_{\rm  vir}$   expressed in terms   of  the critical
density  for closure.    For  the  $\Lambda$CDM  cosmology used   here
$\Delta_{\rm vir} \simeq 101$ (e.g., Bryan \& Norman 1998).
 
When fitting the rotation curves we take adiabatic contraction (Barnes
\& White  1984;  Blumenthal \etal 1986; Flores   \etal  1993) and beam
smearing into account (see BS01 for details).  The best fit values for
$c_{\rm vir}$ and  $V_{\rm vir}$ are listed in Table~\ref{tab:results}
together  with  the inferred   values  for $f_{\rm  disk}$ and $f_{\rm
gas}$. Note that the values of  $c_{\rm vir}$ and $V_{\rm vir}$ differ
somewhat from the values of $c$ and $V_{200}$  quoted in BS01 owing to
the different definitions.

\subsection{Angular momentum distributions}
\label{sec:amd}
 
After finding the mass model that best  fits the observed $\Sigma_{\rm
HI}(r)$  and  $V_{\rm  rot}(r)$  we  determine  $V_c(r)$,  $L(r)$, and
$M_{\rm HI}(r)$  on a linear radial  grid  between $r=0$ and $r=r_{\rm
max}$.  Here $V_c(r)$  is the total  circular  velocity at $r$  of the
best fit model, $L(r)$ is the total $R$-band luminosity inside $r$, as
computed from  the thick exponential,  $M_{\rm HI}(r)$ is the total HI
mass   inside  $r$ computed   from  the best   fit  HI surface density
distribution, and $r_{\rm max}$ is the maximum radius  out to which HI
is detected. We define the total disk mass inside radius $r$ as
\begin{equation}
\label{mbar}
M_{\rm disk}(r) = \Upsilon_{R} \cdot L(r) + 1.3 \cdot M_{\rm HI}(r).
\end{equation}
We thus assume that the  stellar mass-to-light ratio is constant  with
radius, and that the total gas mass is simply  $1.3$ times the HI mass
to  take account of  Helium.   Any contribution from molecular  and/or
ionized gas is therefore considered  negligible.  The distribution  of
specific   angular  momentum is    simply  given  by  $m(j)  =  M_{\rm
disk}(r)/M_{\rm disk}(r_{\rm max})$ with $j=r \, V_c(r)$.
\begin{figure*}
\centerline{\psfig{figure=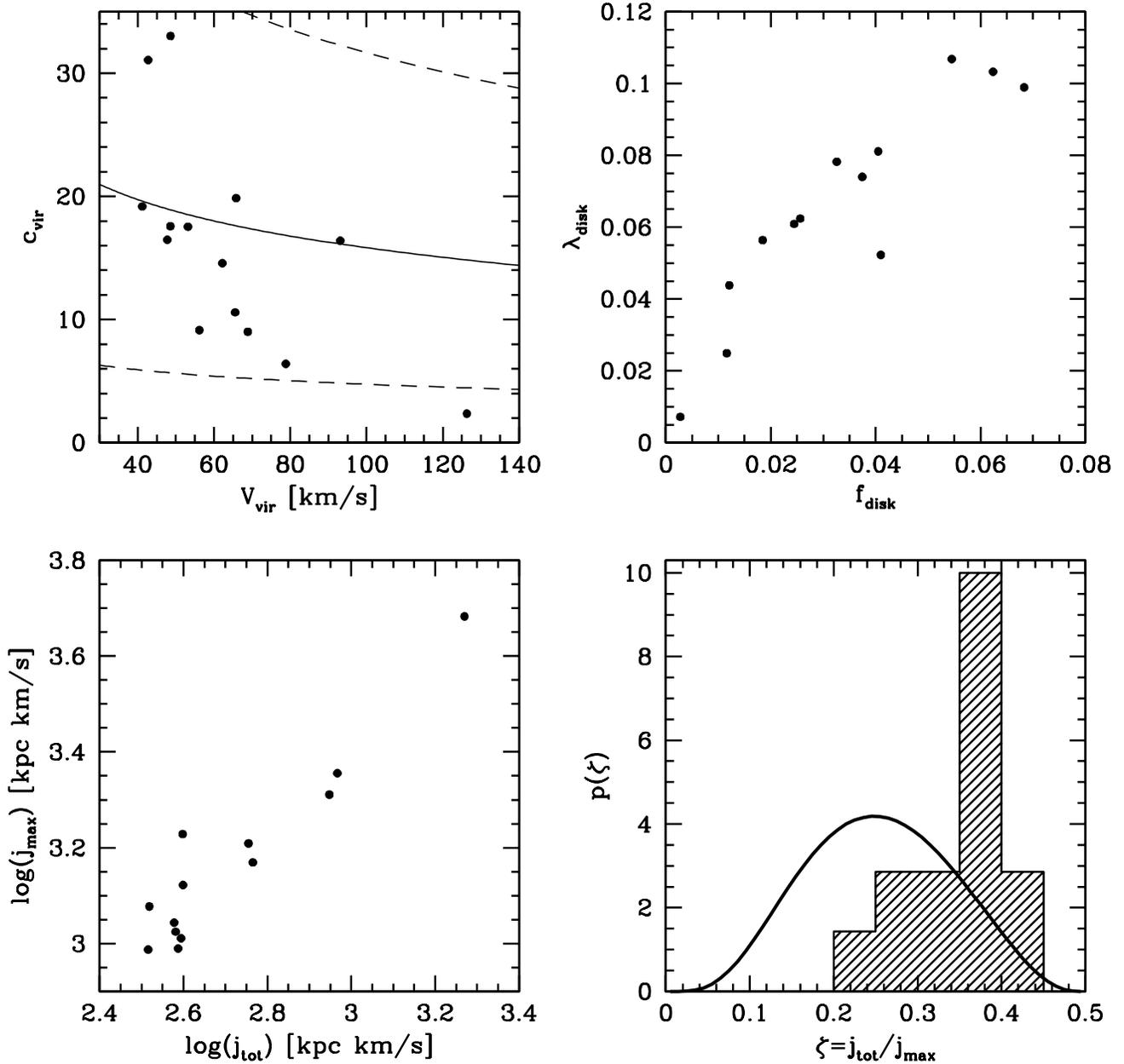,width=\hdsize}}
\caption{The upper left panel plots $c_{\rm vir}$ versus $V_{\rm vir}$
for  the  mass models with   $\Upsilon_R=1.0\MLsun$  that best fit the
observed rotation curves. The solid and dashed lines correspond to the
median and $2 \sigma$ limits of the  distribution of $c_{\rm vir}$ for
the $\Lambda$CDM cosmology adopted here  (computed using the model  of
Bullock  \etal 2001).  Although  the dwarf galaxies  reveal a somewhat
steeper   decline of $c_{\rm   vir}$  with  increasing $V_{\rm  vir}$,
reflecting the degeneracies in the rotation curve fitting, the overall
distribution of halo   concentrations is in  reasonable agreement with
the $\Lambda$CDM cosmology (see BS01 for  a more detailed discussion).
The  upper  right     panel reveals  a   narrow  correlation   between
$\lambda_{\rm bar}$  (determined  from  the  AMDs) and $f_{\rm   bar}$
(inferred from the  best fit mass  models).  The implications of  this
correlation for the formation of disk galaxies are discussed in detail
in the text.   The lower left  panel plots ${\rm log}(j_{\rm max})$ of
the disk versus ${\rm   log}(j_{\rm  tot})$.  The strong   correlation
found implies a narrow  distribution  of $\zeta = j_{\rm   tot}/j_{\rm
max}$, which is   shown in the  lower right  panel (shaded histogram).
The thick  solid curve corresponds  to the distribution  of $\zeta$ of
dark  matter haloes in a  $\Lambda$CDM  cosmology, and is broader  and
offset to lower  values of $\zeta$.  This  reiterates that the AMDs of
the  disks in our sample are  different from the  AMDs  of dark matter
haloes      found   by  B00      (see    also  Figures~\ref{fig:allmj}
and~\ref{fig:allprob}).}
\label{fig:corr}
\end{figure*}

In  Figure~\ref{fig:allmj} we  plot  $m(j)$ as  function  of $j/j_{\rm
max}$ for all the galaxies in our sample.  Results are shown for three
different values of the $R$-band mass-to-light ratio: $\Upsilon_R = 0$
(dotted    lines),  $\Upsilon_R  = 1.0     \MLsun$ (solid lines),  and
$\Upsilon_R =  2.0  \MLsun$ (dashed lines).   For  comparison, we also
plot    the   specific      angular     momentum    distribution    of
equation~(\ref{mb00}) for three  values of $\mu$: $1.06$,  $1.25$, and
$2.0$ (thin solid lines).   These distributions outline the 90 percent
interval   of  the   AMDs        of   $\Lambda$CDM haloes         (see
Section~\ref{sec:theory}).   As  is immediately apparent,  the AMDs of
the disks are clearly different from those of $\Lambda$CDM haloes.

From the  AMDs  thus obtained we   compute the total  specific angular
momentum of the  disk,  $j_{\rm tot}$,  (using  equation~[\ref{zeta}]),
which we parameterize by
\begin{equation}
\label{lambar}
\lambda_{\rm disk} = {j_{\rm tot} \over \sqrt{2}  \, r_{\rm vir} V_{\rm
vir}} \, {\cal G}(1.0,c_{\rm vir}).
\end{equation}
The values  of $j_{\rm tot}$   and $\lambda_{\rm disk}$  are listed in
Table~\ref{tab:results}.  In Figure~\ref{fig:histo} we plot histograms
of the normalized   probability distributions of $\lambda_{\rm  disk}$
for  all  three  values   of   $\Upsilon_{R}$.  For   comparison   the
probability     distribution  of   equation~(\ref{spindistr})     with
$\bar{\lambda} = 0.06$ and   $\sigma_{\lambda}  = 0.5$ is plotted   as
thick solid lines. As can be seen, the spin parameter distributions of
the disks  are fairly  similar to  that   of the  dark  matter haloes,
especially for realistic mass-to-light ratios in the range $1.0 \MLsun
\lta \Upsilon_R \lta 2.0 \MLsun$.  This implies  that, on average, the
{\it total} specific angular momentum of the disks  is similar to that
of the dark matter.  Yet, as we have seen from Figure~\ref{fig:allmj},
their  {\it  distributions} of specific   angular momentum are clearly
distinct  from that of    the  dark matter haloes.   Furthermore,  the
baryonic mass fractions  inferred from  the  best  fit models  to  the
observed rotation curves are significantly  smaller than the Universal
value $f_{\rm bar}$.  All this holds important  clues to the formation
of (dwarf) galaxies, which we discuss in the next section.

\section{The formation of dwarf galaxies}
\label{sec:galform}

In what follows we restrict  ourselves to the  AMDs for $\Upsilon_R  =
1.0 \MLsun$, which is a realistic  value for the stellar mass-to-light
ratio given the typical colors of the dwarfs in our sample (BS01; Bell
\& de Jong 2001). As we address shortly in Section~\ref{sec:uncertain}
below, none of our results  are sensitive to  this particular value of
$\Upsilon_R$.
 
\subsection{Clues from parameter correlations}
\label{sec:corr}

In order to  assess  whether any significant correlations  exist among
the various  parameters listed  in Table~\ref{tab:results}  we compute
the Spearman rank coefficients $r_s$. The resulting correlation matrix
is listed  in Table~\ref{tab:corr}.  The only significant correlations
are those between $V_{\rm vir}$ and $c_{\rm  vir}$ (the probability of
obtaining $r_s = -0.701$ under the null hypothesis that no correlation
exists  is $p_s =    5.2 \times 10^{-3}$),  and between  $\lambda_{\rm
disk}$  and $f_{\rm disk}$, and  $j_{\rm max}$ and $j_{\rm tot}$ (both
with  $r_s=0.886$ and $p_s=2.5  \times 10^{-5}$).   We now address the
meaning of  each  of these  three correlations,  which  are plotted in
Figure~\ref{fig:corr}.
\begin{table}
\begin{minipage}{\hssize}
\caption{Correlation matrix.}
\label{tab:corr}
\begin{tabular}{rrrrrrr}
 & $j_{\rm tot}$ & $j_{\rm max}$ & $f_{\rm disk}$ &
 $f_{\rm gas}$ & $V_{\rm vir}$ & $c_{\rm vir}$ \\
$\lambda_{\rm disk}$ & $0.398$ & $0.231$ & $0.886$ & $0.407$ & $-0.543$ & $0.464$ \\
$j_{\rm tot}$ & $...$ & $0.886$ & $0.538$ & $-0.055$ & $0.332$ & $-0.262$ \\
$j_{\rm max}$ & $...$ & $...$ & $0.363$ & $-0.024$ & $0.516$ & $-0.270$ \\ 
$f_{\rm disk}$ & $...$ & $...$ & $...$ & $0.134$ & $-0.411$ & $0.345$ \\
$f_{\rm gas}$ & $...$ & $...$ & $...$ & $...$ & $-0.446$ & $-0.099$ \\
$V_{\rm vir}$ & $...$ & $...$ & $...$ & $...$ & $...$ & $-0.701$ \\
\end{tabular}

\medskip

Matrix of Spearman Rank's correlation coefficients, $r_s$, for various
parameters of the dwarf galaxies in  our sample. A value of $r_s=+1.0$
($-1.0$) indicates a  perfect correlation (anti-correlation),  whereas
values  of  $r_s$  near  zero   indicate   that the  parameters    are
uncorrelated.  The   only three  significant correlations    are those
between  ($\lambda_{\rm disk}$,$f_{\rm disk}$), ($j_{\rm tot}$,$j_{\rm
max}$),  and  ($V_{\rm vir}$,$c_{\rm vir}$).  For    all other sets of
parameters  the  values of  $r_s$ obtained    are consistent  with  no
correlation.

\end{minipage}
\end{table}

The decrease   of $c_{\rm vir}$  with increasing  $V_{\rm vir}$ is, to
first order, consistent   with what one   expects for a   $\Lambda$CDM
cosmology.  This  is  indicated by the solid  and  dashed lines in the
upper right panel  of  Figure~\ref{fig:corr}, which correspond to  the
mean     and the $2   \sigma$ limits    of  the  distribution of  halo
concentrations as predicted by the Bullock  \etal (2001) model for the
$\Lambda$CDM   cosmology adopted here.   However,  given the amount of
scatter  expected, it   is surprising  that   we find  such  a  strong
anti-correlation; i.e.,  the slope   of the relation   between $V_{\rm
vir}$ and $c_{\rm vir}$ of the dwarf  galaxies in our sample seems too
steep.   This most likely reflects  degeneracies in the rotation curve
fitting: a small under-(over)estimate of $V_{\rm vir}$ can result in a
relatively large over-(under)estimate of $c_{\rm vir}$ (see BS01 for a
more detailed discussion).

The strong correlation between $j_{\rm tot}$ and $j_{\rm max}$ signals
a small dispersion in  the distribution of  $\zeta$.  This is shown in
the lower right panel of  Figure~\ref{fig:corr}. The hatched histogram
corresponds to  the distribution in  $\zeta$ for the 14 dwarf galaxies
in our sample.  We   can   compare this to   the  $\zeta$-distribution
expected for   dark matter haloes   by converting the  distribution in
$\mu$ found by B00 using the correspondence  between $\mu$ and $\zeta$
(equation~[\ref{zetab00}]).  The resulting   probability  distribution
$p(\zeta)$ is plotted as a thick  solid line.  Comparing $p(\zeta)$ of
the dark matter  haloes with that of the  dwarf galaxies one notes two
important differences.   First of   all, the latter   is significantly
narrower  than the   former  (which explains  the  strong  correlation
between $j_{\rm tot}$ and  $j_{\rm max}$).  Secondly, the mean $\zeta$
of the dwarf  galaxies is larger than that  of the dark matter haloes.
Low values of $\zeta$  imply AMDs with  a  long tail to high  specific
angular  momentum, and such AMDs   are apparently underrepresented  in
disk      galaxies compared  to   dark     matter   haloes (but    see
Section~\ref{sec:disksize} below).

The correlation  between $\lambda_{\rm disk}$  and  $f_{\rm disk}$, in
the  sense  that dwarf  galaxies with   a   higher inferred  disk mass
fraction   have  a higher specific  angular    momentum, is remarkably
strong.  Based on   the same set  of  data, but using a  less detailed
computation of $\lambda_{\rm   disk}$ based on the  stellar  component
only, a similarly strong  correlation between $\lambda_{\rm disk}$ and
$f_{\rm disk}$ was recently found by  Burkert (2000b).  At first sight
such correlation is consistent with  an inside-out formation scenario:
the baryons  that  make up stars  and  cold gas have  cooled from  the
inside out  to  form the  galaxies.  Since  most   angular momentum is
contained  in the outermost mass shells,  which  cool latest, one thus
expects  a   correlation as  seen.   However,   the problem  with this
interpretation is that  if only a small fraction  of the  baryons have
cooled, one would  expect the total  specific angular  momentum of the
disks to be significantly smaller than that  of the dark matter.  Yet,
the distribution of $\lambda_{\rm disk}$ is comparable to that of dark
matter haloes,  whereas $f_{\rm  disk}$ is significantly  smaller than
$f_{\rm bar}$.  Apparently,   galactic disks in low-mass systems  form
only  out  of a small   fraction of the total   baryonic mass, but yet
manage to draw most of the  available angular momentum.  This puzzling
aspect has recently  also been noted by  Navarro  \& Steinmetz (2000),
using simple scaling relations of disk galaxies.
\begin{figure*}
\centerline{\psfig{figure=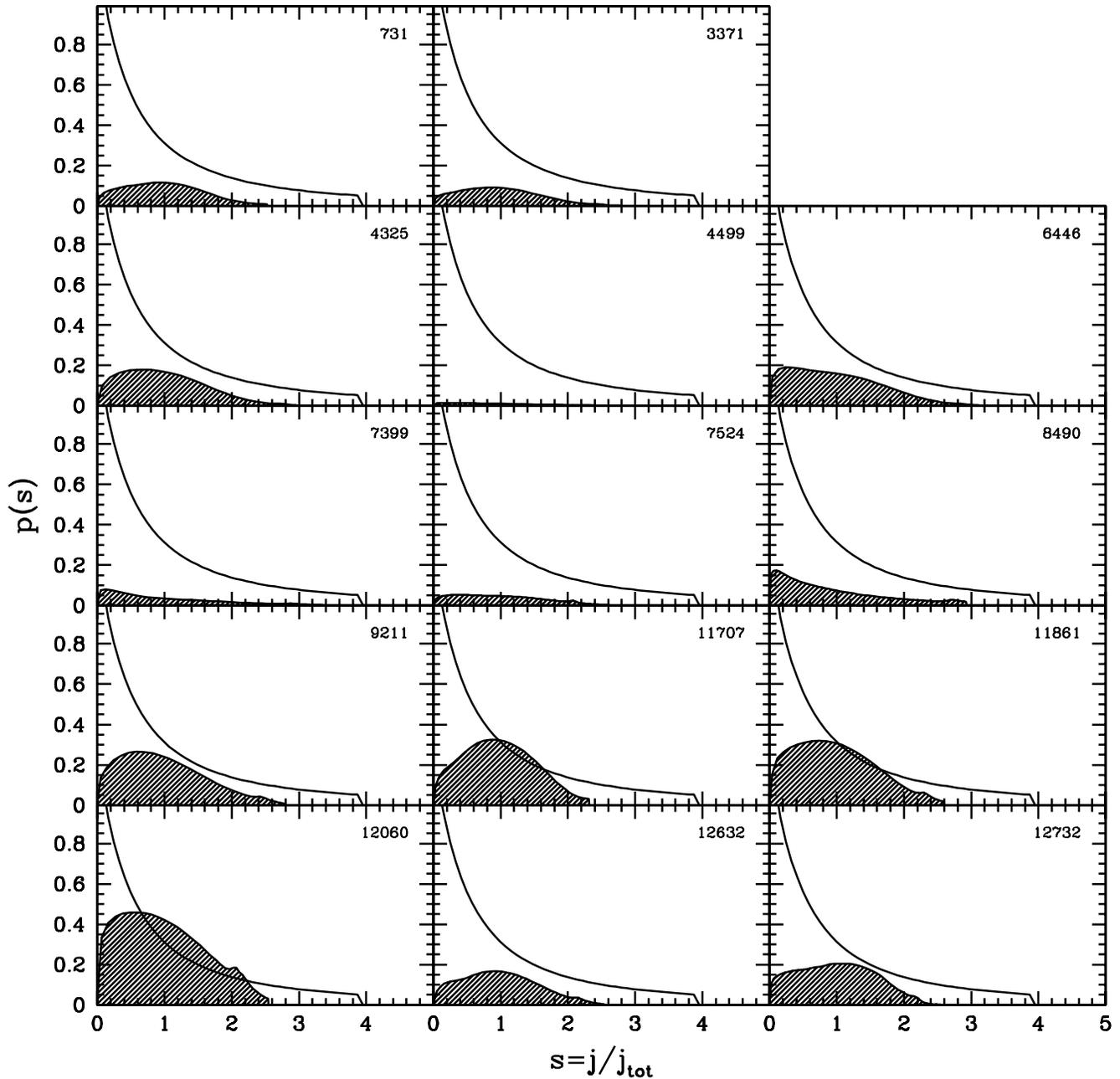,width=\hdsize}}
\caption{The shaded areas  indicate  the AMDs  $p(s)$  of the 14  disk
galaxies in our sample, normalized to $f_{\rm disk}/f_{\rm bar}$.  For
comparison we plot   $p(s)$ of  equation~(\ref{pb00}) with  $\mu=1.25$
(normalized to unity), and which represents the  median of the AMDs of
$\Lambda$CDM haloes.   Under  the  standard  assumption  that   baryons
conserve their specific angular   momentum the difference between  the
two distributions reflects the AMD of  the baryonic matter that is not
incorporated in the disk. Note that it  is preferentially the baryonic
matter with  both the highest and  the lowest angular momentum that is
absent in the disks.}
\label{fig:allprob}
\end{figure*}

\subsection{Challenges for the standard model of disk formation}
\label{sec:model}

In the   standard picture, disks   form out of   baryons that cool and
conserve their specific angular momentum.  Since both baryons and dark
matter experience the same tidal forces,  it is generally assumed that
both have the same angular momentum distribution. Therefore, initially
the  AMD of  the baryonic   component,  with mass $f_{\rm bar}  M_{\rm
vir}$, can be parameterized by equation~(\ref{pb00}).  A comparison of
this distribution  with the probability  distributions $p(s)$  of real
galaxies can  be used  to test  this  standard picture and/or  provide
useful   insights regarding the  details   of galaxy formation.   Such
comparison is  only  valid if the  total  baryonic mass  and  the disk
material  have the same $j_{\rm tot}$   (after all $s=j/j_{\rm tot}$).
The fact that the distribution of  $\lambda_{\rm disk}$ for our sample
of  dwarf galaxies  is  consistent  with $p(\lambda)$  of  dark matter
haloes   (cf.   Figure~\ref{fig:histo}) shows   that this  is  a valid
assumption to make, at least in a statistical sense.

In   Figure~\ref{fig:allprob}  we    plot the  distributions   $p(s)$,
normalized  to $f_{\rm disk}/f_{\rm  bar}$,  for  all galaxies in  our
sample   (hatched areas).   In     addition    we plot  $p(s)$      of
equation~(\ref{pb00}) with $\mu=1.25$ (the median  value found by B00)
and  normalized to unity (solid   lines).   This  distribution is   to
represent the AMD of  the {\it total} baryonic  mass.  Then, under the
standard  assumption  that baryons   conserve  their  specific angular
momentum, the difference  between the two distributions describes  the
AMD of the baryonic material that did not make it into the disk.

The first characteristic to note from Figure~\ref{fig:allprob} is that
the area under the hatched curves is much  smaller than that under the
solid lines, reflecting the fact that  $f_{\rm disk}$ is significantly
smaller than the  Universal baryon fraction  $f_{\rm bar}$.  Secondly,
in all  cases  $p(s)$ of the total   baryonic matter extends   to much
higher specific  angular momenta.    This  indicates that  its   ratio
$j_{\rm max}/j_{\rm tot}$ is  larger  than that  of the disk  material
(cf.  lower right  panel of  Figure~\ref{fig:corr}), and implies  that
none   of the baryonic  material   with the  highest specific  angular
momentum has made it into  the disk component. Finally, the difference
between $p(s)$ of the total baryonic mass and that of the disk mass is
largest for small  $s$.

We  can thus conclude that if  the assumptions made, and which reflect
our standard picture for the formation  of disk galaxies, are correct,
the combined  effects of cooling, feedback, and   stripping have to be
such that: (i)  only a small fraction of  the available baryons end up
in the  disk  component,  (ii)  the total specific    angular momentum
material of the disk is comparable to that of the total baryonic mass,
(iii) none of the highest  specific angular momentum material makes it
into the disk, and (iv) there is a  tendency to preferentially prevent
low angular momentum material from ending up in the disk.
\begin{figure*}
\centerline{\psfig{figure=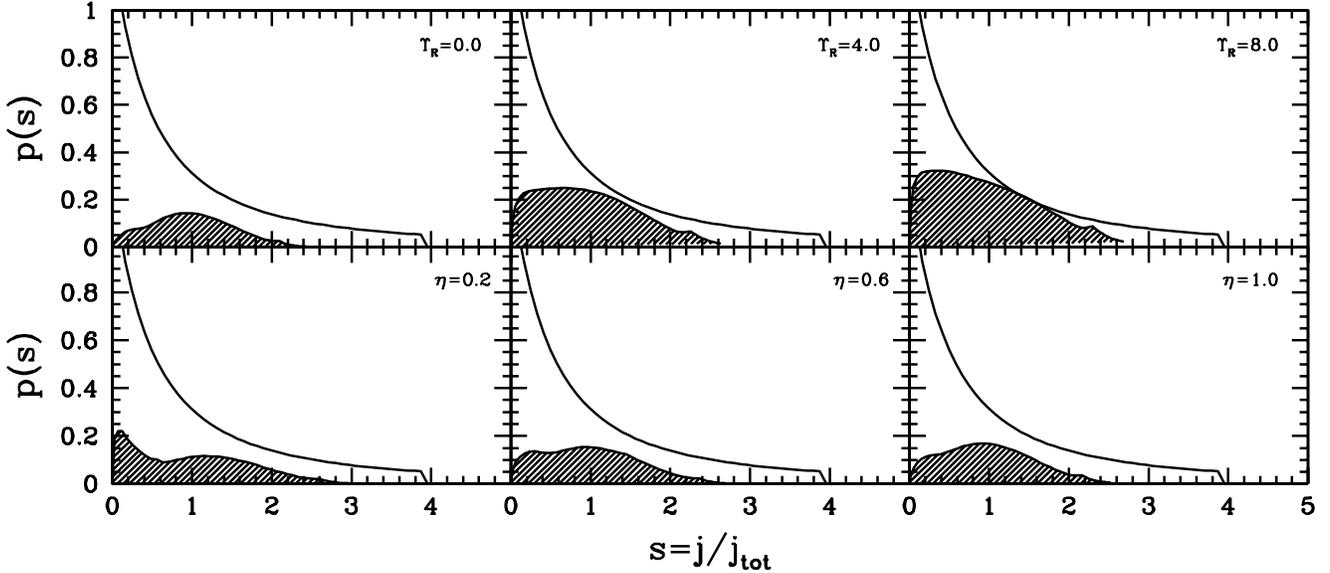,width=\hdsize}}
\caption{The influence of  mass-to-light ratio and asymmetric drift on
the AMD of UGC~12632.   The upper panels  plot $p(s)$ of UGC~12632 for
three  different  values  of  the   stellar  mass-to-light  ratio  (as
indicated in each  panel).  Since the stellar  disk  is more centrally
concentrated  than   the    HI  component,    increasing  $\Upsilon_R$
preferentially increases    the  disk mass  fraction at    low angular
momentum.   However, even  for  $\Upsilon_R=8.0  \MLsun$, the  maximum
mass-to-light ratio allowed by the  rotation curve, a clear deficit at
low $s$  remains.   The lower panels  plot  $p(s)$  but now for  three
different values of $\eta$ (as indicated in  each panel), defined as
the ratio of  the  rotation  velocity  of  the stars to   the circular
velocity ($\eta$ is thus a measure of the  asymmetric drift). Only for
the extreme asymmetric drift corrections with $\eta=0.2$, does the AMD
of UGC~12632 change its characteristic shape. The solid lines are as in
Figure~\ref{fig:allprob}.}
\label{fig:mltest}
\end{figure*}

\subsection{Uncertainties in the angular momentum distributions}
\label{sec:uncertain}

Before drawing any conclusions regarding the formation of (dwarf) disk
galaxies, it is worthwhile to examine some of the uncertainties in the
disk AMDs derived above.

\subsubsection{The Mass of the Disk}
\label{sec:diskmass}

One potential worry  is that we  have missed a significant fraction of
the  actual disk mass.   In the discussion above   we have focussed on
results for $\Upsilon_R = 1.0 \MLsun$,  which, for a Scalo (1986) IMF,
yields colors  in good  agreement  with the   data (see discussion  in
BS01).   However, internal  dust extinction  could   have reddened the
galaxies and/or  a large  fraction of brown  dwarfs could  be present.
Furthermore,   a central   concentration of  molecular   gas could  be
present,  which  would add  low   angular  momentum material.    These
uncertainties in   the disk's mass    distribution can be   modeled by
varying $\Upsilon_R$, either as  a constant (mimicking for example the
presence of  brown dwarfs),  or as  function of  radius (mimicking for
example   the   presence of  molecular      gas).    The freedom    in
$\Upsilon_R(R)$ is limited by the requirement that  the mass model has
to yield  a reasonable fit  to the observed  rotation  curve.  We have
performed  extensive tests with   varying $\Upsilon_R(R)$.  Within the
restrictions imposed by the rotation curves  we find that, although it
is possible  to bring $p(s)$ of the  disk in somewhat better agreement
with that of the total baryonic mass,  the four characteristics listed
above    remain.    This  is  illustrated    in  the  upper panels  of
Figure~\ref{fig:mltest} where we plot   $p(s)$ of UGC~12632  for three
values  of $\Upsilon_R$ (taken to  be constant with radius).  Even for
$\Upsilon_R =  8.0 \MLsun$, which  is the  maximum mass-to-light ratio
allowed by the  observed rotation curve,  there is  still a pronounced
deficit  of low-angular momentum material  in the disk compared to the
total baryonic mass component.

\subsubsection{The Size of the Disk}
\label{sec:disksize}

The maximum  specific angular momentum of the  disk, $j_{\rm max}$, is
directly proportional to the size of  the disk $r_{\rm max}$ (rotation
curves   are  generally flat   at      large radii).  As stated     in
Section~\ref{sec:amd},  we assume that  $r_{\rm  max}$ is equal to the
radius out to  which HI is  detected.  Typically, these radii coincide
with HI column densities of the order of $\sim 10^{19} {\rm cm}^{-2}$.
This is close to the column density below which one expects the gas to
be  ionized by the  cosmic background flux  of ionizing photons (e.g.,
Sunyaev 1969;  Maloney  1993).  Therefore, it might  well  be that the
actual gas disk extends significantly beyond the $r_{\rm max}$ adopted
here. If we make the assumption that this gas follows the same surface
density profile as the HI we can compute the total disk mass missed by
integrating equation~(\ref{sbHI}) out  to infinity, and comparing that
to the   total gas mass inside $r_{\rm   max}$. We find   that for all
galaxies in our sample the gas at radii beyond  $r_{\rm max}$ does not
contribute  more  than $0.5$  percent of  the  total  disk mass. Thus,
whereas the uncertain extent  of the disk  can imply values of $j_{\rm
max}$ that  are significantly underestimated,  this does not influence
our estimates  of $j_{\rm tot}$.  For instance,  if the actual size of
the  disks is about  $1.4$  times $r_{\rm max}$,   the mean of $\zeta$
would be in  much  better agreement with   the expected mean  for dark
matter haloes (lower right panel of Figure~\ref{fig:corr}).  Also, the
thick  curves  in Figure~\ref{fig:allmj} would  shift  to the right by
about the same factor, bringing them in somewhat better agreement with
the   AMDs     of    B00.         However,     Figures~\ref{fig:histo}
and~\ref{fig:allprob} would remain virtually unchanged, and except for
aspect (iii), the main problems outlined above thus remain.

\subsubsection{Asymmetric Drift}
\label{sec:asymdrift}

Another problem with the AMDs derived in Section~\ref{sec:angmomdistr}
is related to  the fact that we  have  assumed that the  stars move on
purely  circular orbits.  In  reality, however,  stars have a non-zero
asymmetric drift, and we have thus  overestimated the angular momentum
of  the  stellar component.  In    order to assess   the importance of
asymmetric  drift we  perform the following   test.  We define a  free
parameter $\eta$ that describes the  (constant) ratio between the true
rotation    velocity of    the   stars   and    the local     circular
velocity\footnote{With  this  definition the asymmetric drift velocity
is  given  by $v_a(r) = (1-\eta)   \, V_c(r)$.}.  In  the  lower three
panels of  Figure~\ref{fig:mltest}   we plot $p(s)$  of UGC~12632  for
$\Upsilon_R=1.0 \MLsun$ and three different  values of $\eta$:  $0.2$,
$0.6$, and $1.0$ (i.e., no asymmetric drift). Only for $\eta \lta 0.3$
do we find  a discernibly  different  AMD.  Typical asymmetric  drifts
correspond to values of $\eta$ much closer  to unity, and we therefore
conclude that our   assumption that $\eta=1.0$ does  not significantly
influence our  results. This can also be  understood directly from the
high  gas  mass   fractions found   for   the dwarf  galaxies   in our
sample. With  $0.4 \lta f_{\rm gas} \lta  0.9$ the disk mass, and thus
the  AMD, of our  dwarf galaxies is generally  dominated by the gas. A
small shift of the angular momentum distribution of the stars relative
to the gas therefore does not significantly influence the total AMD.

\subsubsection{Adiabatic Contraction}
\label{sec:adcorr}

In the mass modelling used to fit the observed rotation curves we have
made the assumption  that the density distribution of  the dark matter
halo  is modified  by  adiabatic contraction.   Since the  assumptions
underlying  the method used  (i.e., sphericity,  adiabatic invariance)
are not  necessarily accurate it  is wortwhile to examine  what effect
this adiabatic  contraction has on our  results. To test  this we have
reanalyzed all data without  any correction for adiabatic contraction.
The effects,  however, are negligible.  Typically one  finds values of
$c_{\rm vir}$ that are slightly larger (the dark matter halo now needs
to  be  more centrally  concentrated  in  order  to fit  the  observed
rotation curve),  but the actual AMDs remain  virtually unchanged. One
parameter that is  directly influenced by the change  in $c_{\rm vir}$
is the spin  parameter $\lambda$ (equation~[{\ref{lamjtot}]), but even
here the differences are small (less than 5 percent). 

Thus,  despite   considerable freedom   in  the  stellar mass-to-light
ratios, the sizes of the disks, and  the asymmetric drift corrections,
we conclude  that the  characteristics of  the  AMDs of disks outlined
above    are    robust.   As     for    the   parameters   listed   in
Table~\ref{tab:results}:  $f_{\rm     gas}$, $\lambda_{\rm  disk}$ and
$j_{\rm tot}$  are very robust,   $j_{\rm max}$ may  be  significantly
underestimated, and the uncertainties on $c_{\rm vir}$, $V_{\rm vir}$,
and  $f_{\rm  disk}$ are strongly  correlated   and discussed in  more
detail in BS01.   Typically,  lowering  $V_{\rm vir}$ implies   larger
$c_{\rm    vir}$       and    $f_{\rm    disk}$.       This,   through
equation~(\ref{lamjtot}) results  in a  slightly larger  $\lambda_{\rm
disk}$.   As long  as  the quality of  the rotation  curve fit remains
similar,  the  total   specific  angular  momentum, however,   remains
virtually   unchanged. This     implies    that of     the  four  main
characteristics listed in Section~\ref{sec:model} above, only (iii) is
somewhat questionable as it is directly related to the assumed size of
the disk.

\subsection{Cooling, stripping, feedback \& viscosity}
\label{sec:barphysics}

If cooling in small mass haloes is very  inefficient, it might explain
aspects (i) and  (iii).  After all, cooling is  an inside out process,
and the highest angular momentum material resides  in the outskirts of
the haloes.  However,  as already mentioned in Section~\ref{sec:corr},
this picture is     inconsistent   with (ii), as   one   would  expect
$\lambda_{\rm disk}$ to be much smaller than the spin parameter of the
dark matter haloes, and it does not explain aspect (iv).  Furthermore,
cooling  is   supposed to be  extremely  efficient  in  dense low mass
haloes, and one typically expects the  vast majority of the baryons in
dwarf  galaxies to have cooled by  the present  time.  Tidal stripping
has virtually  the same effects as  inefficient  cooling; stripping is
most  efficient  near  the outskirts  of the  haloes,  and thus  might
explain  both (i)  and (iii).  However, as  with cooling, stripping is
inconsistent with both (ii) and (iv).

The problem with feedback as the dominant cause  for the AMDs observed
is   that one would  expect  a  relatively  strong correlation between
$f_{\rm  disk}$ and  $V_{\rm vir}$.  After    all, the efficiency  for
blowing a galactic wind should be inversely proportional to the square
of the  galaxy's escape velocity,  which  in turn  is proportional  to
$V_{\rm  vir}$.  However, the  correlation between  $f_{\rm disk}$ and
$V_{\rm vir}$ has a Spearman rank coefficient of $r_s=-0.411$ and thus
seems to   indicate  an   {\it anti}-correlation (although     with  a
probability of $p_s = 0.14$ this is not significant).  Furthermore, as
shown by van den Bosch  (2001), a simple  model for supernovae induced
galactic winds does  not reveal any  tendency to  preferentially expel
the low angular momentum material as required.

One possibility is that one  of the standard  assumptions is wrong and
that baryons do   not   conserve  their specific  angular    momentum.
However, given that disks have the  same distribution of $j_{\rm tot}$
as the   dark  matter haloes implies   that one   can only consider  a
scenario in which   the specific angular  momentum  of the baryons  is
redistributed.   However,  the natural mechanism for  angular momentum
redistribution, viscosity, transports low angular momentum inwards and
high angular  momentum  material  outwards.  Therefore,  any viscosity
present in the disk  will only aggravate  the disagreement at  low and
high $s$ between the AMDs of the disk and the total baryonic matter.

\subsection{The nature of dark matter}
\label{sec:nature}

One interesting interpretation of  the  large differences between  the
AMDs of  disks and cold dark  matter haloes found   might be that dark
matter is not cold.   Recently, several studies  have focussed on warm
and     self-interacting  dark   matter    (hereafter WDM   and  SIDM,
respectively) scenarios,  and have  shown  that  this results in  dark
matter haloes with constant  density cores and with less  substructure
(e.g., Spergel    \&  Steinhardt   2000;   Burkert  2000a;  Col\'{i}n,
Avila-Reese  \& Valenzuela 2000; Bode,  Ostriker \& Turok 2000; Dav\'e
\etal 2001; Kochanek  \& White 2000; Moore  \etal  2000; Yoshida \etal
2000a,b).  Not only might this  solve the angular momentum catastrophe
mentioned  in Section~\ref{sec:intro} (Sommer-Larsen  \& Dolgov 2001),
it  might  also alleviate  the  potential problems  with  the rotation
curves of dwarf and low surface brightness galaxies (Flores \& Primack
1994; Moore 1994; Burkert 1995; McGaugh \& de Blok 1998; van den Bosch
\etal 2000; van den  Bosch \& Swaters 2001)  and with the abundance of
satellite galaxies (Kauffmann, White  \& Guiderdoni 1993;  Moore \etal
1999; Klypin \etal  1999).  It would be  worthwhile to examine to what
extent these differences  in halo  structure result  in AMDs that  are
different from those of CDM haloes.   For instance, Moore \etal (2000)
have pointed out  that   the  central  regions  of SIDM   haloes   are
rotationally supported,  implying AMDs with  less low angular momentum
material  compared to CDM haloes..   Detailed high resolution $N$-body
simulations  of structure formation  in  WDM and SIDM cosmologies  are
required to  test whether a modification of  the nature of dark matter
brings the AMDs of disks and dark matter haloes in better agreement.

\subsection{Decoupling the baryons from the dark matter}
\label{sec:decouple}

Probably the  most  likely conclusion  from  our  results is that  the
standard assumption that baryons and dark  matter have the same AMD is
incorrect.  Since dark    and baryonic  matter   should in   principle
experience the same cosmological torques, we  thus need a mechanism to
somehow decouple  the baryonic  mass component from  that  of the dark
matter.  One such  mechanism,  suggested by  Hogan (1979), is  Compton
drag  on the background radiation.  This  causes the  gas to be frozen
into   the comoving frame such  that  it experiences no tidal torques.
However, this mechanism is only  efficient at very high redshifts, and
will cause the baryons to have less specific angular momentum than the
dark  matter. This  is inconsistent  with  what is required, since, as
shown   by Burkert (2000b),  the   strong correlation between  $f_{\rm
disk}$ and $\lambda_{\rm disk}$  found  requires a mechanism that  can
actually   spin-up  the  baryons relative to     the dark matter.  

A more plausible mechanism might be to  resort to feedback from early
structure formation which stirs  up the baryonic component  resulting
in a density distribution that is quite distinct from that of the dark
matter.  Subsequent  torques will then be  different  for the two mass
components,  causing   a    decoupling  of  their   angular   momentum
distributions. Furthermore, it is important to realize that because of
the dissipative  nature of    the baryonic matter,   angular  momentum
redistribution in merging systems will be quite different for baryonic
and dark matter. High resolution hydro-dynamical simulations should be
useful to investigate whether the AMDs of baryons  and dark matter are
similar (as usually assumed), or whether the processes mentioned above
cause  a decoupling that bring  the AMDs of  the baryonic component in
better agreement with those of disk galaxies.

\section{Summary}
\label{sec:summ}

In the standard picture of structure formation the angular momentum of
protogalaxies originates from  cosmological  torques.  Since dark  and
baryonic matter experience the same  tidal forces, it is expected that
both mass  components end  up with the  same  distribution of specific
angular  momentum.  In a  recent paper Bullock \etal (2000) determined
the AMDs  of dark matter haloes  in a $\Lambda$CDM  cosmology.  If our
picture of disk formation is correct, we thus expect  disks to be born
out of similar distributions.  A comparison of  the AMDs of disks with
those of dark  matter haloes thus  yields important  insights into the
formation mechanism of galaxies, and may  be used to test our standard
picture of disk formation.

In this paper  we  have computed the   AMDs of a   sample of 14  dwarf
galaxies with observed HI rotation curves.  These rotation curves have
been analyzed in detail by van den Bosch  \& Swaters (2001), who found
them  to  be in  good  agreement  with   $\Lambda$CDM predictions.   A
comparison with the AMDs of dark matter haloes obtained by B00 reveals
the following:

\begin{itemize}

\item The ratios  of disk mass to  total virial  mass, $f_{\rm disk}$,
inferred from mass models fitted to the  observed rotation curves, are
much  smaller than  the  Universal baryon fraction  $f_{\rm  bar}$ for
currently popular cosmologies.  This indicates that large fractions of
baryonic  mass have either been  prevented from cooling,  or have been
removed  from the disk  or    halo by either  feedback  or   stripping
mechanisms.

\item The  distribution of $\lambda_{\rm  disk}$  is in good agreement
with  the  distribution of halo  spin  parameters. This  suggests that
disks  have  the same {\it  total}  specific angular  momentum as dark
matter haloes.

\item  The disk mass fractions  $f_{\rm disk}$ are strongly correlated
with the disk's spin parameter $\lambda_{\rm disk}$.

\item The distribution  of $\zeta =  j_{\rm tot}/j_{\rm max}$ of disks
is narrower than  $p(\zeta)$  of dark matter  haloes, and  with a mean
that is  significantly higher. This implies  that  the AMDs  of haloes
have more extended tails to high specific angular momentum.

\item The normalized angular momentum distributions of (low mass) disk
galaxies are clearly distinct from  those of dark  matter haloes.  The
latter have AMDs that extent  to higher specific angular momentum, and
with much more mass at low angular momentum.

\end{itemize}
Despite uncertainties in the  disk's density distribution,  related to
the  unknown mass-to-light ratio and the  unknown  amount of molecular
gas, these results are robust, and confirm  recent findings of Navarro
\& Steinmetz (2000) that apparently disks form out of a small fraction
of  the available baryonic  mass,  but yet  manage to  tap most of the
available specific angular momentum.

Understanding  these findings  within the  standard framework of  disk
formation is  challenging.  Neither the main   mechanisms that lead to
small values of $f_{\rm disk}$, i.e.,  feedback, stripping or cooling,
nor viscous  processes that  redistribute specific  angular  momentum,
seem able to provide  a meaningful explanation.  Somehow these results
seem  to imply that   the baryonic mass components  out  of which disk
galaxies form through cooling have angular momentum distributions that
are clearly  distinct from  those found by  B00.  This  implies either
that some mechanism decoupled the baryons from  the dark matter during
the early  stages  of the formation  of   galaxies (during which  they
acquire angular  momentum   from cosmological torques),  or  that  the
distributions found by B00 are poor descriptions of the actual AMDs of
dark matter haloes (perhaps reflecting a different  nature of the dark
matter).  It is  clear that  without  a proper understanding of   (the
origin  of) the  angular momentum distribution   of the baryonic  mass
component  of  protogalaxies  our picture  of  the formation   of disk
galaxies is incomplete.


\section*{Acknowledgements}

We are grateful to  Tom Abel, Houjun  Mo, Yi-Peng Jing and Simon White
for stimulating discussions. FB thanks the hospitality  of the MPIA in
Heidelberg during his visit that started the work presented here.


\label{lastpage}

\end{document}